\documentclass[%
 reprint,
superscriptaddress,
 amsmath,
 amssymb,
 aps,
floatfix,
twocolumn
]{revtex4-2}

\usepackage{graphicx}
\usepackage{dcolumn}
\usepackage{bm}
\usepackage[utf8]{inputenc}
\usepackage{tabularx}
\usepackage[dvipsnames]{xcolor}
\usepackage{units}
\usepackage[euler]{textgreek}
\usepackage{upgreek}
\usepackage{derivative}
\usepackage{amsmath}
\usepackage{physics}
\usepackage{amssymb}
\usepackage{hyperref}
\hypersetup{colorlinks=true,linkcolor=blue!50!black,urlcolor=blue!50!black,citecolor=blue!50!black}
\usepackage{colortbl}
\usepackage{svg}
\usepackage{romannum}
\usepackage{booktabs}
\usepackage{array}
\begin{document}

\preprint{APS}

\title{Deep-underground dark matter search with a COSINUS detector prototype}

\author{G.~Angloher}
\affiliation{Max-Planck-Institut f\"ur Physik, 80805 M\"unchen - Germany}

\author{M.R.~Bharadwaj}
\email{Corresponding author: mukund@mpp.mpg.de}
\affiliation{Max-Planck-Institut f\"ur Physik, 80805 M\"unchen - Germany}

\author{I.~Dafinei}
\affiliation{Gran Sasso Science Institute, 67100 L'Aquila - Italy}
\affiliation{INFN - Sezione di Roma, 00185 Roma - Italy}

\author{N.~Di~Marco}
\affiliation{Gran Sasso Science Institute, 67100 L'Aquila - Italy}
\affiliation{INFN - Laboratori Nazionali del Gran Sasso, 67010 Assergi - Italy}

\author{L.~Einfalt}
\email{Corresponding author: leonie.einfalt@oeaw.ac.at}
\affiliation{Institut f\"ur Hochenergiephysik der \"Osterreichischen Akademie der Wissenschaften, 1050 Wien - Austria}
\affiliation{Atominstitut, Technische Universit\"at Wien, 1020 Wien - Austria}

\author{F.~Ferroni}
\affiliation{Gran Sasso Science Institute, 67100 L'Aquila - Italy}
\affiliation{INFN - Sezione di Roma, 00185 Roma - Italy}

\author{S.~Fichtinger}
\affiliation{Institut f\"ur Hochenergiephysik der \"Osterreichischen Akademie der Wissenschaften, 1050 Wien - Austria}

\author{A.~Filipponi}
\affiliation{INFN - Laboratori Nazionali del Gran Sasso, 67010 Assergi - Italy}
\affiliation{Dipartimento di Scienze Fisiche e Chimiche, Universit\`a degli Studi dell'Aquila, 67100 L'Aquila - Italy}

\author{T.~Frank}
\affiliation{Max-Planck-Institut f\"ur Physik, 80805 M\"unchen - Germany}

\author{M.~Friedl}
\affiliation{Institut f\"ur Hochenergiephysik der \"Osterreichischen Akademie der Wissenschaften, 1050 Wien - Austria}

\author{A.~Fuss}
\affiliation{Institut f\"ur Hochenergiephysik der \"Osterreichischen Akademie der Wissenschaften, 1050 Wien - Austria}
\affiliation{Atominstitut, Technische Universit\"at Wien, 1020 Wien - Austria}

\author{Z.~Ge}
\affiliation{SICCAS - Shanghai Institute of Ceramics, 200050 Shanghai - P.R.C}

\author{M.~Heikinheimo}
\affiliation{Helsinki Institute of Physics, University of Helsinki, 00014 Helsinki - Finland}

\author{M.N. ~Hughes}
\affiliation{Max-Planck-Institut f\"ur Physik, 80805 M\"unchen - Germany}

\author{K.~Huitu}
\affiliation{Helsinki Institute of Physics, University of Helsinki, 00014 Helsinki - Finland}

\author{M.~Kellermann}
\email{Corresponding author: moritz.kellermann@mpp.mpg.de}
\affiliation{Max-Planck-Institut f\"ur Physik, 80805 M\"unchen - Germany}

\author{R.~Maji}
\affiliation{Institut f\"ur Hochenergiephysik der \"Osterreichischen Akademie der Wissenschaften, 1050 Wien - Austria}
\affiliation{Atominstitut, Technische Universit\"at Wien, 1020 Wien - Austria}

\author{M.~Mancuso}
\affiliation{Max-Planck-Institut f\"ur Physik, 80805 M\"unchen - Germany}

\author{L.~Pagnanini}
\affiliation{Gran Sasso Science Institute, 67100 L'Aquila - Italy}
\affiliation{INFN - Laboratori Nazionali del Gran Sasso, 67010 Assergi - Italy}

\author{F.~Petricca}
\affiliation{Max-Planck-Institut f\"ur Physik, 80805 M\"unchen - Germany}

\author{S.~Pirro}
\affiliation{INFN - Laboratori Nazionali del Gran Sasso, 67010 Assergi - Italy}

\author{F.~Pr\"obst}
\affiliation{Max-Planck-Institut f\"ur Physik, 80805 M\"unchen - Germany}

\author{G.~Profeta}
\affiliation{INFN - Laboratori Nazionali del Gran Sasso, 67010 Assergi - Italy} 
\affiliation{Dipartimento di Scienze Fisiche e Chimiche, Universit\`a degli Studi dell'Aquila, 67100 L'Aquila - Italy}

\author{A.~Puiu}
\affiliation{Gran Sasso Science Institute, 67100 L'Aquila - Italy}
\affiliation{INFN - Laboratori Nazionali del Gran Sasso, 67010 Assergi - Italy}

\author{F.~Reindl}
\affiliation{Institut f\"ur Hochenergiephysik der \"Osterreichischen Akademie der Wissenschaften, 1050 Wien - Austria}
\affiliation{Atominstitut, Technische Universit\"at Wien, 1020 Wien - Austria}

\author{K.~Sch\"affner}
\affiliation{Max-Planck-Institut f\"ur Physik, 80805 M\"unchen - Germany}

\author{J.~Schieck}
\affiliation{Institut f\"ur Hochenergiephysik der \"Osterreichischen Akademie der Wissenschaften, 1050 Wien - Austria}
\affiliation{Atominstitut, Technische Universit\"at Wien, 1020 Wien - Austria}

\author{D.~Schmiedmayer}
\affiliation{Institut f\"ur Hochenergiephysik der \"Osterreichischen Akademie der Wissenschaften, 1050 Wien - Austria}
\affiliation{Atominstitut, Technische Universit\"at Wien, 1020 Wien - Austria}

\author{C.~Schwertner}
\affiliation{Institut f\"ur Hochenergiephysik der \"Osterreichischen Akademie der Wissenschaften, 1050 Wien - Austria}
\affiliation{Atominstitut, Technische Universit\"at Wien, 1020 Wien - Austria}

\author{K.~Shera}
\affiliation{Max-Planck-Institut f\"ur Physik, 80805 M\"unchen - Germany}

\author{M.~Stahlberg}
\affiliation{Max-Planck-Institut f\"ur Physik, 80805 M\"unchen - Germany}

\author{A.~Stendahl}
\affiliation{Helsinki Institute of Physics, University of Helsinki, 00014 Helsinki - Finland}

\author{M.~Stukel}
\affiliation{Gran Sasso Science Institute, 67100 L'Aquila - Italy}
\affiliation{INFN - Laboratori Nazionali del Gran Sasso, 67010 Assergi - Italy}

\author{C.~Tresca}
\affiliation{INFN - Laboratori Nazionali del Gran Sasso, 67010 Assergi - Italy} 
\affiliation{Dipartimento di Scienze Fisiche e Chimiche, Universit\`a degli Studi dell'Aquila, 67100 L'Aquila - Italy}

\author{F.~Wagner}
\affiliation{Institut f\"ur Hochenergiephysik der \"Osterreichischen Akademie der Wissenschaften, 1050 Wien - Austria}

\author{S.~Yue}
\affiliation{SICCAS - Shanghai Institute of Ceramics, 200050 Shanghai - P.R.C}

\author{V.~Zema}
\affiliation{Max-Planck-Institut f\"ur Physik, 80805 M\"unchen - Germany}

\author{Y.~Zhu}
\affiliation{SICCAS - Shanghai Institute of Ceramics, 200050 Shanghai - P.R.C}

\collaboration{The COSINUS Collaboration}
\noaffiliation

\begin{abstract}
Sodium iodide (NaI) based cryogenic scintillating calorimeters using quantum sensors for signal read out have shown promising first results towards a model-independent test of the annually modulating signal detected by the DAMA/LIBRA dark matter experiment.
The COSINUS collaboration has previously reported on the first above-ground measurements using a dual channel readout of phonons and light based on transition edge sensors (TESs) that allows for particle discrimination on an event-by-event basis.
In this letter, we outline the first underground measurement of a NaI cryogenic calorimeter read out via the novel remoTES scheme. A \unit[3.67]{g} NaI absorber with an improved silicon light detector design was operated at the Laboratori Nazionali del Gran Sasso, Italy.

A significant improvement in the discrimination power of $e^-$/$\gamma$-events to nuclear recoils was observed with a five-fold improvement in the nuclear recoil baseline resolution, achieving $\sigma$ = \unit[441]{eV}. 
Furthermore, we present a limit on the spin-independent dark-matter nucleon elastic scattering cross-section achieving a sensitivity of $\mathcal{O}$(pb) with an exposure of only \unit[11.6]{g\,d}.
\end{abstract}

\maketitle
\section{Introduction}
Dark matter (DM) detection remains one of the crucial experimental challenges of present-day particle physics and cosmology. 
Direct DM-searches aim to detect potential DM particle candidates via interactions in earth-bound detectors, but most direct detection efforts have returned null results thus far.
In contrast, the DAMA/LIBRA experiment observes an annual modulation of the interaction rate in sodium iodide (NaI) crystals with a statistical significance of 13.7$\sigma$ \cite{bernabei2022recent}, a characteristic signature expected from such DM candidates.
Yet, the DAMA signal could so far neither be confirmed by any other experiment, nor be explained by a non-DM origin.

With the advent of new detection techniques, the direct detection community has made significant progress in pushing the sensitivity for DM-nucleus interactions.
Today, cryogenic detectors (CRESST-III \cite{PhysRevD.100.102002}, EDELWEISS \cite{PhysRevD.99.082003}, SuperCDMS \cite{PhysRevLett.120.061802}) compete for the lowest thresholds while highest exposures are reached using liquid noble gas detectors (LZ \cite{PhysRevC.104.065501}, Panda-X \cite{Ning2023}, XENONnT \cite{PhysRevD.103.063028}). 
Despite numerous experiments covering the parameter space compatible with the DAMA/LIBRA signal under standard assumptions \cite{PhyStat}, the origin of the modulated signal remains to be conclusively proven. 
To reduce the systematic uncertainties of comparing different materials, multiple experiments using the same target material as DAMA/LIBRA are under construction (PICO-LON \cite{Fushimi_2016}, SABRE \cite{refId0}) or already taking data (ANAIS \cite{PhysRevD.103.102005}, COSINE-100 \cite{PhysRevD.106.052005}, DM-Ice \cite{PhysRevD.95.032006}).

The COSINUS (Cryogenic Observatory for SIgnatures seen in Next-generation Underground Searches) experiment utilizes NaI-based scintillating calorimeters operated at cryogenic temperatures and read out using transition edge sensors (TESs).
By measuring the scintillation light as well as the phonon signal of atomic recoils, the COSINUS experiment is the only NaI-based experiment able to discriminate the interaction type on an event-by-event basis.
Furthermore, it is possible to perform in-situ measurements of the quenching factor, setting the signal region for recoils off Na and I.

With one year of data taking, first results on a possible nuclear recoil origin of the DAMA/LIBRA signal are anticipated, while \unit[1000]{kg\,d} will provide a complete model-independent cross check \cite{Angloher_2016,Kahlhoefer_2018}.

In 2021, COSINUS developed the remoTES detector design to build low-threshold detectors for materials so far not operable as cryogenic detectors in a reliable and reproducible way for future mass production. A follow-up study devoted to the application of this design to NaI, a highly challenging absorber crystal for cryogenic searches,  was successfully carried out, observing particle discrimination on an event-by-event basis in a NaI-experiment for the first time \cite{ANGLOHER2023167532, IDM}.
The study reported here describes an improved detector design operated in an underground cryostat at the Laboratori Nazionali del Gran Sasso (LNGS), yielding the best energy resolution achieved by a NaI-based detector for nuclear recoils.
Section \ref{module} describes the detector module design, which includes a NaI phonon channel with remoTES readout as well as a silicon (Si) light channel with TES readout.
In sections \ref{analysis} and \ref{limit}, the collected data are analyzed, and we present the resulting DM-nucleon scattering cross-section limit for this measurement.

\section{Detector module}\label{module}

\begin{figure}[ht]
\includegraphics[width=.48\textwidth]{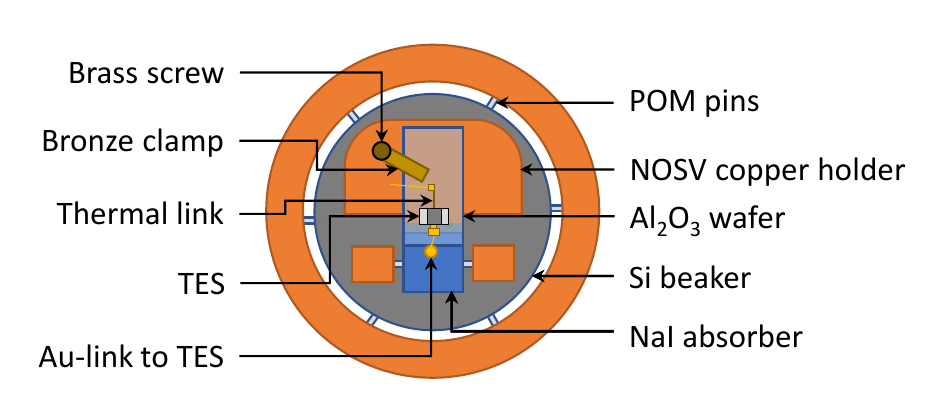}
\caption{\label{fig:detector_bird}Top view of the detector module.}
\end{figure}

\begin{figure}[ht]
\includegraphics[width=.48\textwidth]{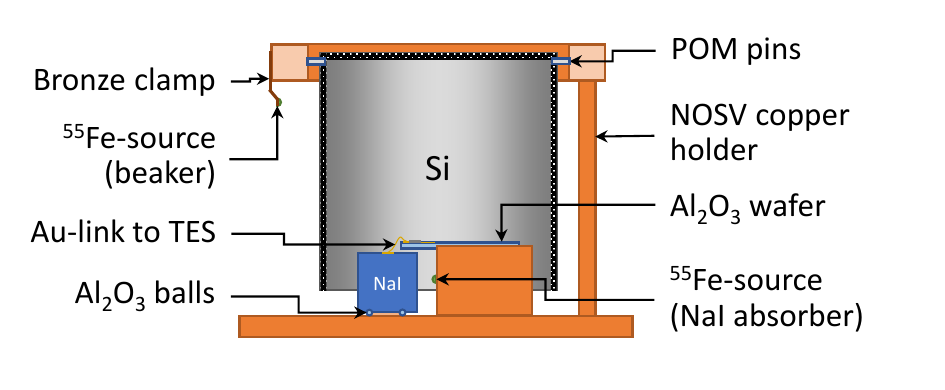}
\caption{\label{fig:detector_xsec}Cross-sectional view of the detector module.}
\end{figure}

The detector module consists of two independent channels: 
a NaI absorber operated as a cryogenic calorimeter (phonon channel) and a light absorber to detect the corresponding scintillation photons generated by a particle interaction in NaI (light channel). 
A schematic breakdown of the module is depicted in Fig.~\ref{fig:detector_bird} and \ref{fig:detector_xsec}. 

The phonon channel consists of a \unit[3.67]{g} NaI crystal mounted in a holder fabricated from electrolytic tough pitch (NOSV) copper (Cu) \cite{aurubis}.
The NaI crystal was produced by Shanghai Institute for Ceramics, China (SICCAS) with a modified Bridgman technique as described in \cite{siccas} using ``Astro-Grade" powder procured from Merck group \cite{merckgroup}. 
ICP-MS (Inductively Coupled Plasma Mass Spectrometry) measurements performed at LNGS \cite{LNGS} revealed an internal contamination at a level of \unit[6-22]{ppb} for $^{40}$K, $<$\unit[1]{ppb} for $^{208}$Th and $^{238}$U respectively.
The crystal used for this particular measurement had a thallium (Tl)-dopant level of \unit[$730 \pm 73$]{ppm}.  

The crystal rests on a trio of $\textrm{Al}_2\textrm{O}_3$ balls to thermally insulate it from the Cu holder. 
Two additional support tips made out of polyoxymethylene (POM) fix the crystal's position. 
An $^{55}$Fe X-ray source with an activity of \unit[0.11]{Bq} is taped onto the Cu holder such that it irradiated one of the faces of the crystal. 
The resultant K$_{\alpha}$ and K$_{\beta}$ lines are used to calibrate the detector response in the offline analysis.
The remoTES scheme, described and implemented in \cite{pyle, ANGLOHER2023167532}, utilizes a gold (Au) link for signal read out via the TES from particle interactions within the absorber. For the present study, this link consists of an Au-pad adhered to NaI, with a thin Au-wire connecting the pad to the TES.
All components of the phonon channel and their properties are shown in Tab. \ref{tab:dimensions1}.
The TES consists of a tungsten-based superconducting thin-film evaporated onto a $\textrm{Al}_2\textrm{O}_3$ wafer using infrastructure and technology of the CRESST group at the Max-Planck Institute for Physics (MPP) in Munich, Germany.
A gold stripe (thermal link) connecting the TES to the thermal bath is used for weak thermal coupling (\unit[65]{$\Omega$} at room temperature) to slowly dissipate heat. 
An ohmic heater film was deposited onto the wafer to adjust the TES-temperature to the optimal operation point. Externally injected "test pulses" at regular intervals via the heater serve to precisely measure the detector response over its entire dynamic range and to monitor potential changes with time.

 \begin{table}[!htb]
 \renewcommand{\arraystretch}{1.4}
     \centering
     \begin{tabular}{ll}
         \textbf{Component}&  \textbf{~Properties} \\ \toprule
          NaI absorber~          &  ~Volume: (10x10x10) mm$^3$ \\ \hline
        Au-link~                 &  ~Au-pad on NaI \\
                                 &  ~~~~~Area: \unit[1.77]{mm$^2$} \\
                                 &  ~~~~~Thickness: \unit[1]{$\upmu$m} \\
                                 &  ~~~~~Glue: EPO-TEK 301-2 \cite{epotek} \\ 
                                 &  ~Au-wire \\
                                 &  ~~~~~Length: $\sim$ \unit[10]{mm}\\
                                 &  ~~~~~Diameter: \unit[17]{$\upmu$m} \\ \hline
        Al$_2$O$_3$ wafer~       &  ~Volume: (10x20x1) mm$^3$ \\ 
        W-TES on wafer~          &  ~Area: (100x400)~$\upmu$m$^2$ \\
                                 &  ~Thickness: \unit[156]{nm} \\ 
                                 &  ~T$_C$: \unit[28]{mK} \\
        Heater on wafer~         &  ~Area: \unit[(200x150)]{$\upmu$m$^2$} \\
                                 &  ~Thickness: \unit[100]{nm} gold\\
        \bottomrule 
    \end{tabular}
    \caption{Properties of the phonon channel.}
    \label{tab:dimensions1}
\end{table}
To collect the scintillation light, a beaker-shaped Si crystal with a mass of \unit[15.38]{g} was used. 
It was mounted on a separate Cu frame with the help of six POM tips (applying even pressure from all sides).
An $^{55}$Fe X-ray source with an activity of \unit[3.3]{mBq} was taped onto the Cu holder to irradiate the beaker. 
A tungsten TES was deposited directly on the Si beaker.
To efficiently collect athermal phonons and deliver their energy, it is flanked by two superconducting aluminium phonon collectors \cite{phonon_collectors}.
A thermal link (\unit[15.8]{$\Omega$} at room temperature) and a separate heater were also deposited on the Si beaker, similar to the scheme described for the phonon channel TES. 
All components of the light channel and their features are shown in Tab. \ref{tab:dimensions2}.

\begin{table}[!htb]
\renewcommand{\arraystretch}{1.4}
   \centering
   \begin{tabular}{ll}
   \textbf{Component} &  \textbf{~Properties}\\\hline
        Si-absorber~           &  ~Hollow cylinder\\
                               &  ~Height: \unit[40]{mm}\\ 
                               &  ~Outer diameter: \unit[40]{mm}\\
                               &  ~Thickness: \unit[1]{mm}  \\ 
       W-TES on Si~            &  ~Area: (100x400)~$\upmu$m$^2$ \\
                                &  ~Thickness: \unit[80]{nm} \\ 
                                &  ~T$_C$: \unit[28]{mK} \\
       Heater on Si~         &  ~Area: \unit[(200x150)]{$\upmu$m$^2$} \\
                                &  ~Thickness: \unit[100]{nm} gold\\ \hline
        
   \end{tabular}
   \caption{Properties of the light channel.}
   \label{tab:dimensions2}
\end{table}

\begin{figure}[ht]
\includegraphics[width=.4\textwidth]{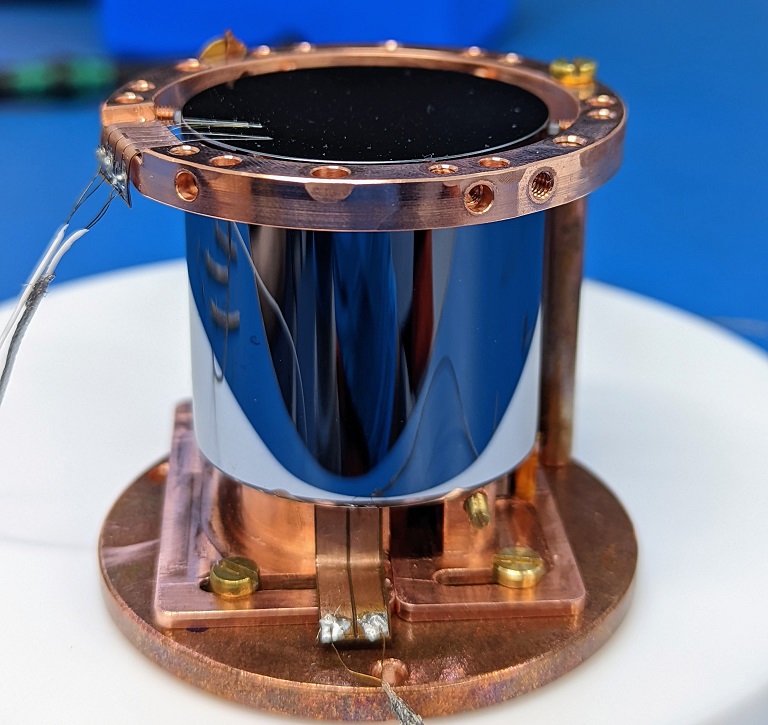}
\caption{\label{fig:photo_module}Photograph of the assembled detector module, consisting of the phonon and light channel.}
\end{figure}

With the help of a Cu pillar, the Si beaker is positioned to enclose the NaI crystal as depicted in Fig.~\ref{fig:detector_xsec}. 
A photograph of the complete, dual channel system is shown in Fig.~\ref{fig:photo_module}. 
The measurement was carried out in a wet dilution refrigerator provided by the CRESST group of MPP.
It is located underground in a side tunnel between hall\,A and hall\,B at LNGS with an overburden of \unit[3600]{m\,w.e.} \cite{overburden}.
A detailed breakdown of the measuring setup is provided in section \Romannum{1} of the supplemental material \cite{supplemental}.
\section{Data analysis} \label{analysis}
The detector signals were read out in parallel with two separate systems at a sampling rate of \unit[50]{kS/s}, a commercial, hardware-triggered DAQ and a custom-made continuous DAQ. 
To set up and stabilize the detectors, the hardware-triggered DAQ was used.

In total about ten days of stream data were taken over the first two weeks of June, 2022. Sixteen hours were measured with an external $^{57}$Co $\gamma$-source (\unit[122]{keV}) with an activity of \unit[430]{Bq} located inside the external lead-shield.
About twenty six hours were measured with an AmBe neutron-source with an activity of \unit[2000]{Bq} located outside the external lead-shield.
Additionally, a simulation of the entire setup was carried out to cross-check the expected neutron interaction rates, and was found to agree well with the measured rate in the energy range of interest.
Additional details about the simulation itself can be found in section \Romannum{3} of the supplemental material \cite{supplemental}.
With a background data-taking period lasting for \unit[76]{h}, an overall exposure of \unit[11.6]{g\,d} was collected.

\subsection{Raw data analysis} \label{lowlevelanalysis}
The data used in the analysis were acquired with the continuous DAQ system, using offline triggering with an optimum filter (OF) trigger on the continuous stream \cite{Gatti:1986cw}. The particle pulse standard events (SE), the noise power spectrum (NPS), and the threshold necessary for the triggering procedure were determined via the hardware triggered data. More details on these components can be found in section \Romannum{2}.A of the supplemental material \cite{supplemental}. 

A set of quality cuts was applied, which aim to discard pulse shapes different from the SE, artifacts caused by interference, as well as noise triggers. The majority of these cuts only affects the phonon channel. The OF amplitude is used to reconstruct the amplitude of the pulses. Moreover, the amplitudes are corrected for small drifts in the detector response over time using the test pulses. 

The energy calibration in both channels is performed by fitting a double Gaussian peak to the $^{55}$Mn K$_\alpha$ (\unit[5.9]{keV}) and K$_\beta$ (\unit[6.5]{keV}) lines of the built-in $^{55}$Fe X-ray sources. As the iron lines are close to threshold in both light and phonon channel, an additional calibration was performed with a $^{57}$Co $\gamma$-source (\unit[122]{keV}). In order to avoid energies outside the linear range of the detector, we limit the region of interest (ROI) to a maximum energy of 200 keV in both channels.

\begin{figure*}[ht]
\includegraphics[width=\textwidth]{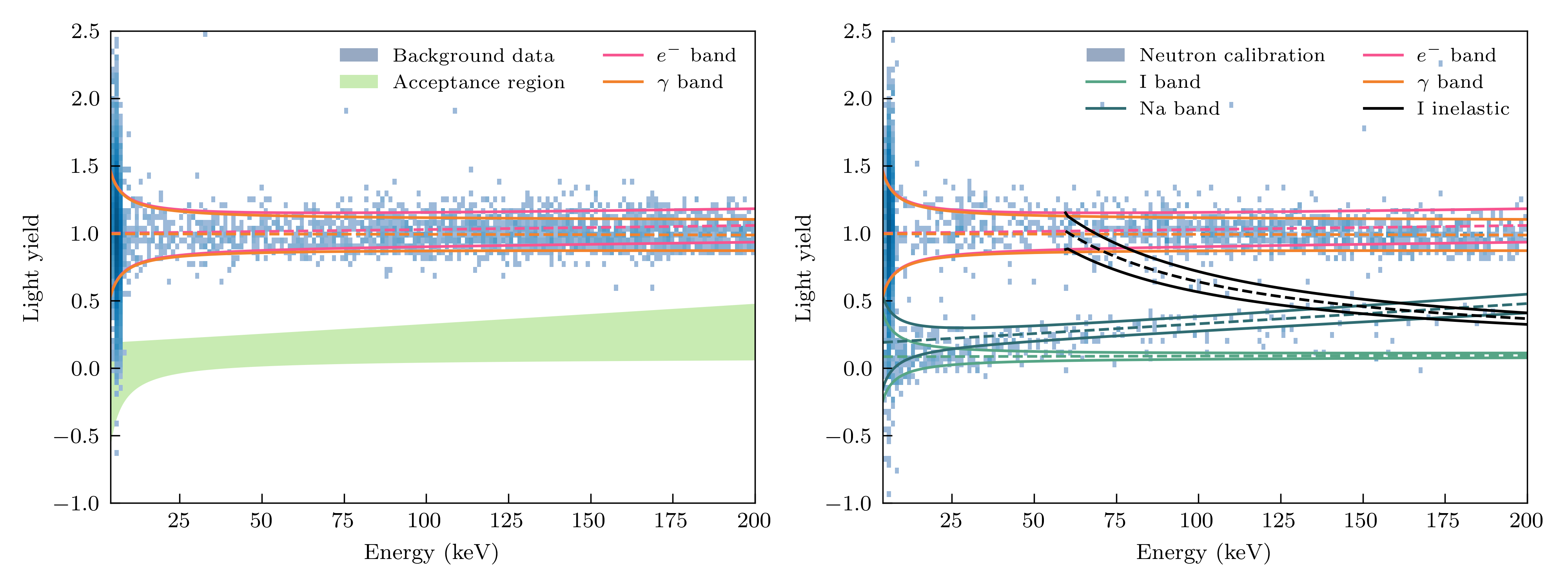}
\caption{\label{fig:bandfit} 2D Histogram of light yield vs.~the total deposited energy for both background (left) and neutron calibration data (right). Both figures show the fit to the $e^-$ (pink) and the $\gamma$ band (orange) as found by the combined likelihood fit. Together with the neutron calibration data we also show the fit results for the nuclear and inelastic recoil bands. The light green shaded region in the left panel marks the acceptance region for DM inference with the Yellin method.}
\end{figure*}
To determine the baseline resolution we superimpose the SE particle template upon cleaned, empty noise traces picked at random times from the full stream. Fitting a Gaussian to the distribution of the filtered amplitudes of these simulated events estimates the resolution as $\sigma_p = 0.3779 \pm 0.0086\,$mV/ $0.441 \pm 0.011\,$keV in the phonon and $\sigma_l = 0.930 \pm 0.021\,$mV/ $0.988 \pm 0.052\,$keV electron equivalent (keV$_{\text{ee}}$) in the light channel \cite{MANCUSO2019492}. The phonon detector threshold of $2.656 \pm 0.041\,$keV was determined from the trigger survival probability (see \Romannum{2}.A of \cite{supplemental}). We evaluate the trigger and cut survival probabilities by applying the whole analysis chain to $\mathcal{O}\left( 10000\right)$ artificial events. The result of this procedure can be found in section \Romannum{2}.A of the supplemental material \cite{supplemental}. For practical purposes of this prototype measurement, in all the following steps, we set the analysis threshold to $E_\text{thr} = 4\,$keV, a value at which the cut efficiency has reached about 50\% of its plateauing value.

\subsection{High-level analysis} \label{highlevelanalysis}

The total deposited energy is calculated using $E=(1-\eta) E_p+\eta E_l$, where $E_p$~($E_l$) is the energy in the phonon (light) channel, respectively. Details on the determination of the parameter $\eta$ can be found in section II of the supplemental material \cite{supplemental}. Following the results of the raw data analysis, the ROI is set to [4, 200] keV. Moreover, we restrict the light yield $LY = E_l/E_p$ to [-10, 10].

The novel feature of COSINUS, compared to other DM searches with NaI target materials, is the combination of phonon and light signal which is used to discriminate between $e^-$/$\gamma$ and nuclear recoil events via their different light yield. Fig.~\ref{fig:bandfit} displays the light yield vs.~energy scatter plot for both background and neutron calibration data. The quenched nuclear recoil events in the neutron calibration are clearly separable from the bulk of the $e^-$ and $\gamma$ recoil events. Under the assumption that DM particles will mainly recoil off nuclei, the acceptance region for the DM analysis (employing Yellin's optimum interval method \cite{Yellin1,Yellin2}) is dependent on the position of the nuclear recoil bands in Fig.~\ref{fig:bandfit}. The positions of the $e^-$, $\gamma$ and nuclear recoil bands are determined by an unbinned likelihood-fit to the whole data set (background and neutron calibration), in both phonon and light energy simultaneously. The parameterization of the energy spectra and energy-dependent light quenching are non-trivial and described in the supplemental material \cite{supplemental}. From the results of the likelihood fit we determined the quenching factors at 10 keV nuclear recoil energy as $QF_{\text{Na}}(10\,\text{keV}) =0.2002\pm 0.0093$ and $QF_{\text{I}}(10\,\text{keV})=0.0825\pm 0.0034$. We want to highlight that the quenching factors are measured intrinsically and specifically for this crystal, such that no systematic uncertainties arise and the small uncertainties stem purely from the fit.

\section{Dark Matter Result} \label{limit}
To give a comparable measure of the detector performance and the impact of the event-by-event discrimination on a DM analysis, we use the background data set to obtain limits on the nucleon-DM spin-independent elastic scattering cross section. The expected DM interaction rate $\odv{R_\text{det}}{E}$ as observed by the detector is characterized by the standard spin-independent scattering model. It includes detector-specific quantities such as the threshold $E_\text{thr}$, the trigger and cut efficiency $\varepsilon(E)$, as well as the probability $p_\text{ACR}(E)$ that a DM recoil event lies within the acceptance region. The resolution of the phonon channel is dominated by the Gaussian baseline noise and, therefore, taken into account by convolution with a Gaussian $Ga$ of width $\sigma_p$:
\renewcommand{\arraystretch}{0.9}
\begin{align}
\begin{split}
    \label{eq:DM_rate}
    \odv{R_\text{det}}{E} \left(E\right)=&\,\,\theta(E-E_\text{thr})\,\varepsilon(E)\,\,p_\text{ACR}(E) \\
    &\int_0^\infty \odv{R_\text{theo}}{E'}(E') Ga(E-E',\sigma_p^2) \dd{E'}
\end{split}
\end{align}

The theoretical model $\odv{R_\text{theo}}{E'}$ is based on the standard assumptions for an isothermal DM halo \cite{PhyStat}. Effects of the nuclear shape for the sodium and iodine nuclei are modeled by the Helm (extended by Lewin and Smith) form factor \cite{FF_Helm,FF_Lewin}, while the low contribution of Tl is conservatively considered negligible in the calculation.

\begin{figure}[t]
\includegraphics[width=0.48\textwidth]{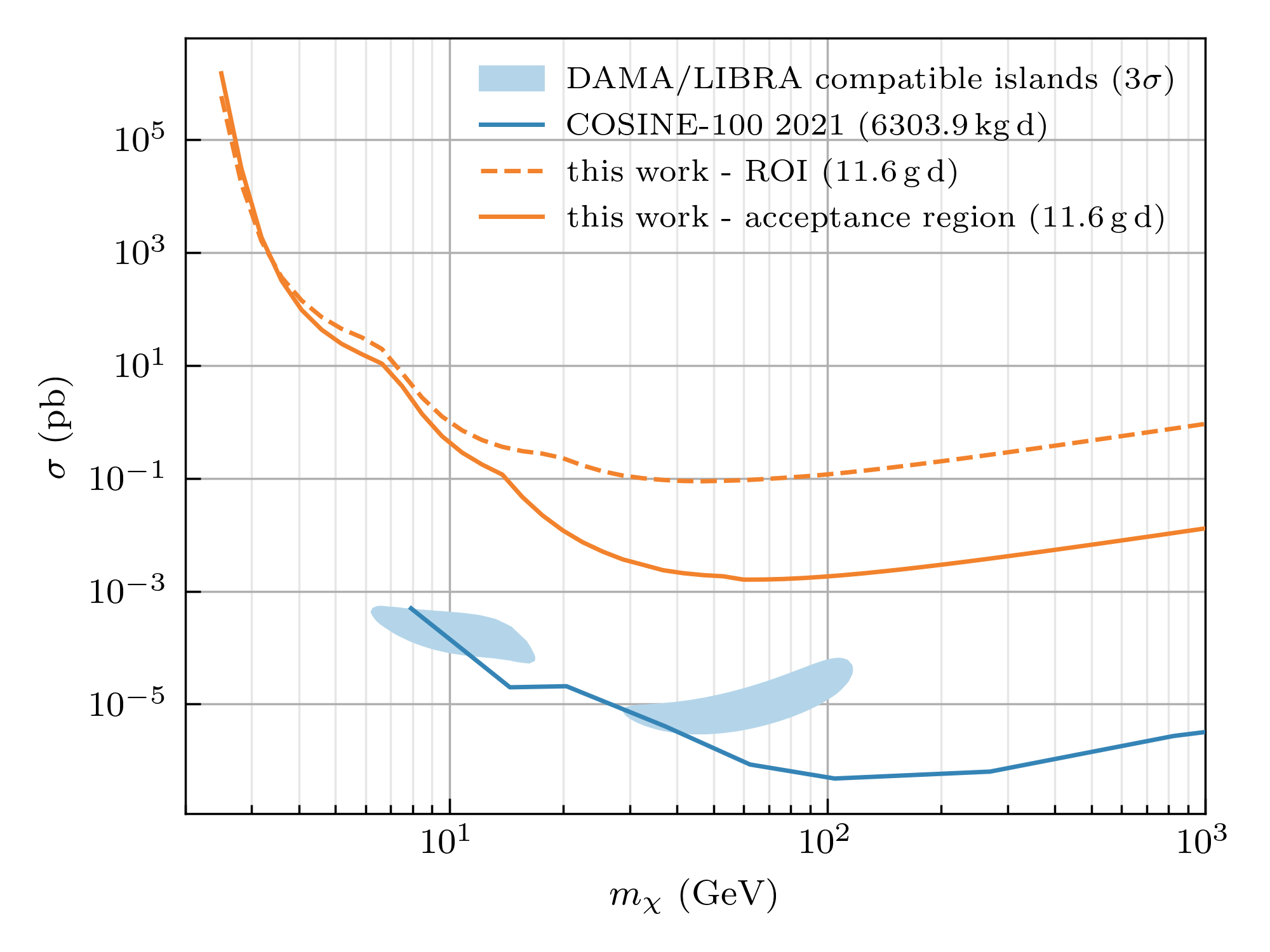}
\caption{\label{fig:limit}90\% confidence level upper limit on the spin-independent, elastic nucleon-DM scattering cross-section in the standard scenario, $\sigma$(pb) as a function of the DM mass, m$_\chi$. The orange lines show the results of this work from a background data set with \unit[11.6]{g\,d} of exposure. The dashed line is the limit achieved considering all events in the ROI; for the solid line only the events in the acceptance region were considered. As a comparison, we show contours compatible with the DAMA/LIBRA result \cite{DAMAislands} and the COSINE-100 result from \unit[6303.9]{kg\,d} exposure \cite{COSINE59days}, a factor $\sim$10$^5$ higher than the current study.}
\end{figure}

Using Yellin's optimum interval method \cite{Yellin1,Yellin2}, we obtain the 90\% confidence level upper limits displayed in Fig.~\ref{fig:limit}. Besides the limit calculated from the acceptance region (solid line), we also show the limits calculated from all data points in the ROI (dashed line). A comparison of the two lines shows, that COSINUS' unique event-by-event discrimination enables to set an up to two orders stricter limit in the standard scenario. We emphasize that the limits from the acceptance region with \unit[11.6]{g\,d} of exposure are only three orders of magnitude less strict than the limits from COSINE-100 with \unit[6303.9]{kg\,d} exposure (blue line) shown for comparison. In addition, we note that the sensitivity of the prototype was limited due to leakage from the $e^-$/$\gamma$ band to the nuclear recoil bands. The majority of these leakage events stem from the $^{55}$Fe calibration source, as well as the overall $e^-$/$\gamma$ background present in the dilution refrigerator. We will have a significant improvement for such measurements in the final low-background facility of COSINUS \cite{facility}.  This is also directly related to the convergence of the limits from the acceptance region (solid orange) and the ROI (dashed orange) at lower DM masses, which correspond to lower energies where the discrimination power is reduced.  Thus, we expect a profile-likelihood ratio approach for the DM analysis, instead of the Yellin method, to further improve the sensitivity \cite{likelihood}.

\section{Conclusion}
In this work we present the results of the first underground operation of a NaI based cryogenic scintillating calorimeter. A baseline resolution of $0.441\,$keV for nuclear recoils was achieved for the phonon channel. Together with the Si-based light channel, the dual channel readout was operated successfully, enabling particle discrimination between $e^-$/$\gamma$ and nuclear recoils on an event-by-event basis. Based on these results, we determine the energy dependent quenching factors for sodium and iodine as observed in the operated crystal to $QF_{\text{Na}}(10\,\text{keV}) =0.2002\pm 0.0093$ and $QF_{\text{I}}(10\,\text{keV})=0.0825\pm 0.0034$.
Furthermore, we give limits on the standard spin-independent, elastic scattering cross section, based on \unit[11.6]{g\,d} exposure of this R\&D run's background data set, demonstrating how our unique background discrimination increases the sensitivity of a COSINUS detector with respect to NaI experiments with single-channel readout.

Future R\&D campaigns will focus on up-scaling the detector design to house undoped NaI crystals of the order of $\mathcal{O}(\unit[10]{g})$, as planned for the final experimental setup of COSINUS, while further lowering the threshold for nuclear recoils.

\begin{acknowledgments}
We would like to thank the CRESST group at the Max Planck Institute for Physics, Munich for allowing us to use their cryogenic test facility at LNGS and detector production infrastructure in Munich, as well as for sharing their cryogenic expertise with us.
We are also grateful to LNGS for their generous support and technical assistance in ensuring a smooth run underground. The offline data analysis was mainly carried out with collaboration-internal software tools \textit{CAT} and \textit{limitless}, and cross checks partially with the open source Python package \textit{CAIT} \cite{wagner2022cait}. Additionally, we would like to thank the MPP and HEPHY mechanical workshop teams for their timely assistance in detector holder fabrication.
This work has been supported by the Austrian Science Fund FWF, stand-alone project AnaCONDa [P 33026-N] and by the Austrian Research Promotion Agency (FFG), project ML4CPD.
\end{acknowledgments}


\bibliography{main_refs.bib}
\end{document}


\preprint{APS}

\title{Supplemental material: Deep-underground dark matter search with a COSINUS detector prototype
}
\author{G.~Angloher}
\affiliation{Max-Planck-Institut f\"ur Physik, 80805 M\"unchen - Germany}

\author{M.R.~Bharadwaj}
\email{Corresponding author: mukund@mpp.mpg.de}
\affiliation{Max-Planck-Institut f\"ur Physik, 80805 M\"unchen - Germany}

\author{I.~Dafinei}
\affiliation{Gran Sasso Science Institute, 67100 L'Aquila - Italy}
\affiliation{INFN - Sezione di Roma, 00185 Roma - Italy}

\author{N.~Di~Marco}
\affiliation{Gran Sasso Science Institute, 67100 L'Aquila - Italy}
\affiliation{INFN - Laboratori Nazionali del Gran Sasso, 67010 Assergi - Italy}

\author{L.~Einfalt}
\email{Corresponding author: leonie.einfalt@oeaw.ac.at}
\affiliation{Institut f\"ur Hochenergiephysik der \"Osterreichischen Akademie der Wissenschaften, 1050 Wien - Austria}
\affiliation{Atominstitut, Technische Universit\"at Wien, 1020 Wien - Austria}

\author{F.~Ferroni}
\affiliation{Gran Sasso Science Institute, 67100 L'Aquila - Italy}
\affiliation{INFN - Sezione di Roma, 00185 Roma - Italy}

\author{S.~Fichtinger}
\affiliation{Institut f\"ur Hochenergiephysik der \"Osterreichischen Akademie der Wissenschaften, 1050 Wien - Austria}

\author{A.~Filipponi}
\affiliation{INFN - Laboratori Nazionali del Gran Sasso, 67010 Assergi - Italy}
\affiliation{Dipartimento di Scienze Fisiche e Chimiche, Universit\`a degli Studi dell'Aquila, 67100 L'Aquila - Italy}

\author{T.~Frank}
\affiliation{Max-Planck-Institut f\"ur Physik, 80805 M\"unchen - Germany}

\author{M.~Friedl}
\affiliation{Institut f\"ur Hochenergiephysik der \"Osterreichischen Akademie der Wissenschaften, 1050 Wien - Austria}

\author{A.~Fuss}
\affiliation{Institut f\"ur Hochenergiephysik der \"Osterreichischen Akademie der Wissenschaften, 1050 Wien - Austria}
\affiliation{Atominstitut, Technische Universit\"at Wien, 1020 Wien - Austria}

\author{Z.~Ge}
\affiliation{SICCAS - Shanghai Institute of Ceramics, 200050 Shanghai - P.R.C}

\author{M.~Heikinheimo}
\affiliation{Helsinki Institute of Physics, University of Helsinki, 00014 Helsinki - Finland}

\author{M.N. ~Hughes}
\affiliation{Max-Planck-Institut f\"ur Physik, 80805 M\"unchen - Germany}

\author{K.~Huitu}
\affiliation{Helsinki Institute of Physics, University of Helsinki, 00014 Helsinki - Finland}

\author{M.~Kellermann}
\email{Corresponding author: moritz.kellermann@mpp.mpg.de}
\affiliation{Max-Planck-Institut f\"ur Physik, 80805 M\"unchen - Germany}

\author{R.~Maji}
\affiliation{Institut f\"ur Hochenergiephysik der \"Osterreichischen Akademie der Wissenschaften, 1050 Wien - Austria}
\affiliation{Atominstitut, Technische Universit\"at Wien, 1020 Wien - Austria}

\author{M.~Mancuso}
\affiliation{Max-Planck-Institut f\"ur Physik, 80805 M\"unchen - Germany}

\author{L.~Pagnanini}
\affiliation{Gran Sasso Science Institute, 67100 L'Aquila - Italy}
\affiliation{INFN - Laboratori Nazionali del Gran Sasso, 67010 Assergi - Italy}

\author{F.~Petricca}
\affiliation{Max-Planck-Institut f\"ur Physik, 80805 M\"unchen - Germany}

\author{S.~Pirro}
\affiliation{INFN - Laboratori Nazionali del Gran Sasso, 67010 Assergi - Italy}

\author{F.~Pr\"obst}
\affiliation{Max-Planck-Institut f\"ur Physik, 80805 M\"unchen - Germany}

\author{G.~Profeta}
\affiliation{INFN - Laboratori Nazionali del Gran Sasso, 67010 Assergi - Italy} 
\affiliation{Dipartimento di Scienze Fisiche e Chimiche, Universit\`a degli Studi dell'Aquila, 67100 L'Aquila - Italy}

\author{A.~Puiu}
\affiliation{Gran Sasso Science Institute, 67100 L'Aquila - Italy}
\affiliation{INFN - Laboratori Nazionali del Gran Sasso, 67010 Assergi - Italy}

\author{F.~Reindl}
\affiliation{Institut f\"ur Hochenergiephysik der \"Osterreichischen Akademie der Wissenschaften, 1050 Wien - Austria}
\affiliation{Atominstitut, Technische Universit\"at Wien, 1020 Wien - Austria}

\author{K.~Sch\"affner}
\affiliation{Max-Planck-Institut f\"ur Physik, 80805 M\"unchen - Germany}

\author{J.~Schieck}
\affiliation{Institut f\"ur Hochenergiephysik der \"Osterreichischen Akademie der Wissenschaften, 1050 Wien - Austria}
\affiliation{Atominstitut, Technische Universit\"at Wien, 1020 Wien - Austria}

\author{D.~Schmiedmayer}
\affiliation{Institut f\"ur Hochenergiephysik der \"Osterreichischen Akademie der Wissenschaften, 1050 Wien - Austria}
\affiliation{Atominstitut, Technische Universit\"at Wien, 1020 Wien - Austria}

\author{C.~Schwertner}
\affiliation{Institut f\"ur Hochenergiephysik der \"Osterreichischen Akademie der Wissenschaften, 1050 Wien - Austria}
\affiliation{Atominstitut, Technische Universit\"at Wien, 1020 Wien - Austria}

\author{K.~Shera}
\affiliation{Max-Planck-Institut f\"ur Physik, 80805 M\"unchen - Germany}

\author{M.~Stahlberg}
\affiliation{Max-Planck-Institut f\"ur Physik, 80805 M\"unchen - Germany}

\author{A.~Stendahl}
\affiliation{Helsinki Institute of Physics, University of Helsinki, 00014 Helsinki - Finland}

\author{M.~Stukel}
\affiliation{Gran Sasso Science Institute, 67100 L'Aquila - Italy}
\affiliation{INFN - Laboratori Nazionali del Gran Sasso, 67010 Assergi - Italy}

\author{C.~Tresca}
\affiliation{INFN - Laboratori Nazionali del Gran Sasso, 67010 Assergi - Italy} 
\affiliation{Dipartimento di Scienze Fisiche e Chimiche, Universit\`a degli Studi dell'Aquila, 67100 L'Aquila - Italy}

\author{F.~Wagner}
\affiliation{Institut f\"ur Hochenergiephysik der \"Osterreichischen Akademie der Wissenschaften, 1050 Wien - Austria}

\author{S.~Yue}
\affiliation{SICCAS - Shanghai Institute of Ceramics, 200050 Shanghai - P.R.C}

\author{V.~Zema}
\affiliation{Max-Planck-Institut f\"ur Physik, 80805 M\"unchen - Germany}

\author{Y.~Zhu}
\affiliation{SICCAS - Shanghai Institute of Ceramics, 200050 Shanghai - P.R.C}

\collaboration{The COSINUS Collaboration}
\noaffiliation

\begin{abstract}
In this supplemental material, we give details on the measurement setup including photographs. Moreover, we provide details on the data cleaning and analysis chain and illustrate these procedures with accompanying figures. A simulation of the expected neutron rate in the phonon channel for the AmBe calibration was performed and compared with the measured rate, the results of which are shown.
\end{abstract}

\maketitle
\section{Measurement setup}\label{module_sup}

\begin{figure}[ht]
\includegraphics[width=.45\textwidth]{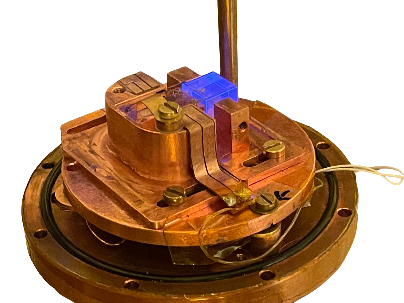}
\caption{\label{fig:photo_NaI}Photograph of the assembled phonon channel read out using the remoTES design. The crystal is irradiated with UV-light to demonstrate its scintillation properties.}
\end{figure}

Final assembly of the complete module took place in a moisture-regulated nitrogen glovebox ($<$ 50 ppm) to avoid degradation of the NaI(Tl) crystal.
Fig.~\ref{fig:photo_NaI} shows a photograph of the assembled phonon detector with a scintillating NaI absorber.

The measurement was carried out at LNGS in a $^3$He/$^4$He-dilution refrigerator of type MINIKELVIN\,400-TOF from Leiden Cryogenics B.V. \cite{leiden} provided by the CRESST group of the Max-Planck Institute for Physics.
Two Superconducting Quantum Interference Devices (SQUIDs) manufactured by Applied Physics Systems (APS) \cite{APS} are used for signal amplification \cite{Karo_PHD}.

\begin{figure}[ht]
\includegraphics[width=0.48\textwidth]{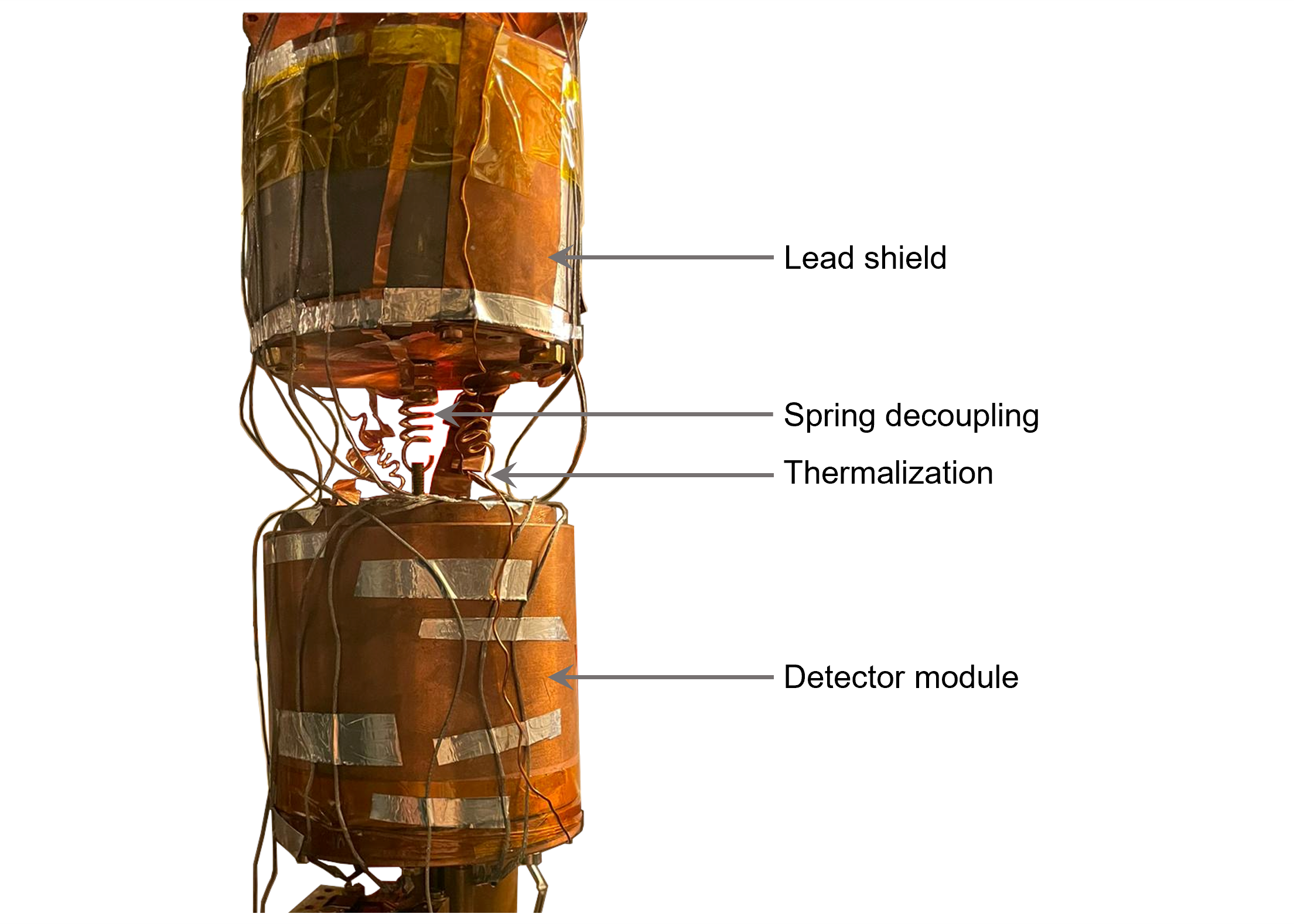}
\caption{\label{fig:mounting}Photograph of the detector module mounted onto the mixing chamber of the dilution refrigerator.}
\end{figure}

The refrigerator is equipped with an external Pb-shield with a thickness of \unit[100]{mm}. A cylindrical internal radiation shield made from low-background lead (Pb) with a diameter of \unit[900]{mm} and a thickness of \unit[100]{mm} is mounted above the physical volume.
To decouple the sensors from vibrations, the detector module was appended on a bronze (CuSn6)-spring with a resonance frequency of a few Hertz.
The thermalization of all parts was ensured by screwed Cu-wire connections to the mixing chamber.
A photograph of the mounting scheme is shown in Fig.~\ref{fig:mounting}.

\section{Data analysis} \label{analysis}
\subsection{Raw data analysis} \label{lowlevelanalysis_sup}
The various components of the offline triggering process are described in detail in this section, starting with the noise power spectrum (NPS). Several hundred empty and cleaned noise traces were collected from the hardware-triggered (HWT) background data, and their respective NPS were averaged. From the same HWT background data, a common set of pulses for both channels is selected from a narrow energy region, summed up, and re-scaled to get a first estimate of the standard event (SE). The SEs are then fitted with the pulse model described in \cite{Zema:2020mkm} to eliminate any remaining noise. In a remoTES-detector we expect different pulse shapes attributed to energy depositions in different detector parts \cite{ANGLOHER2023167532}. As we are interested in recoil events taking place in the NaI crystal, we use only \textit{absorber events} for generating the SEs and in the subsequent analysis steps. The trigger threshold was determined employing the method described in \cite{MANCUSO2019492}, using empty and cleaned noise traces from the HWT data. With the criterion of one noise trigger per \unit[]{kg\,d} exposure, we fix the trigger threshold at 2 mV in the phonon and 6 mV in the light channel. 

\begin{figure}[t]
\includegraphics[width=.48\textwidth]{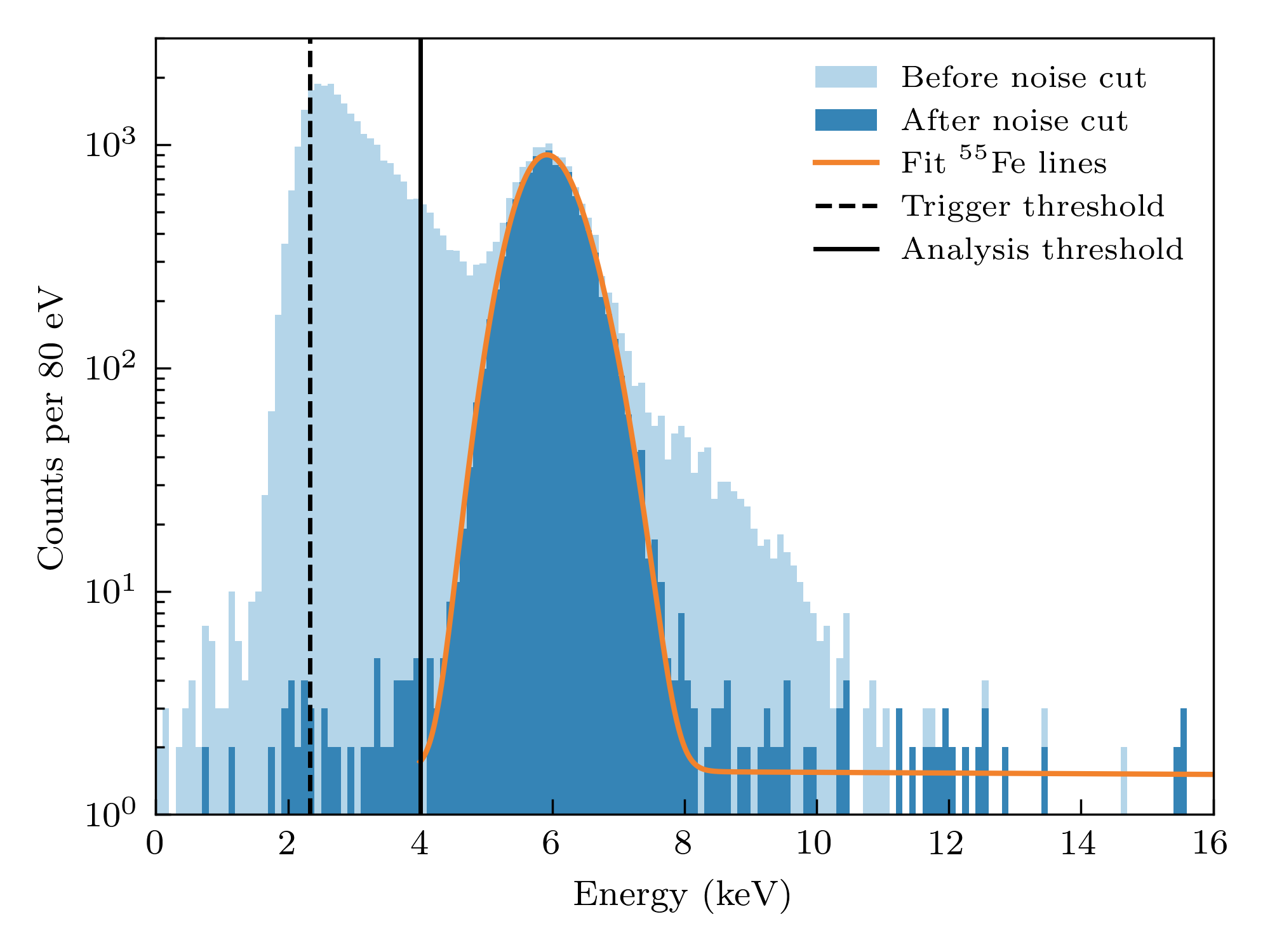}
\caption{\label{fig:noise}The energy spectrum of the background data set below 16 keV before (light blue) and after the noise-leakage removal cut (blue). The black dashed line marks the threshold which was used for the optimum filter trigger, the solid black line the threshold which was used in the subsequent analysis steps. A fit to the the $^{55}$Mn K$_\alpha$ (5.89 keV) and K$_\beta$
(6.49 keV) lines yields the orange line, yielding a resolution of the detector at these energies of $0.450 \pm 0.007\,$keV.}
\end{figure}

As mentioned in the main text, we apply a set of quality cuts to the triggered pulses. We want to describe one cut performed on the phonon channel in more detail. Although the software trigger thresholds determined above should ensure only minimal leakage of noise events over the threshold, there is a significant accumulation of events visible in the phonon channel close to the triggering threshold, shown in Fig.~\ref{fig:noise}. Due to the non-gaussian shape of this noise distribution, we assume that the method in \cite{MANCUSO2019492} is not entirely applicable to our prototype measurement, and the point of one noise trigger per \unit[]{kg\,d} would be at a higher voltage. At low energies, one expects that the reconstruction of a true particle pulse amplitude with an optimum filter should yield the same result as fitting a SE to the pulse. In order to remove the noise leakage and maintain a low analysis threshold, we compare these two amplitudes for the pulses in the phonon channel and disregard any events where the values differ more than 20\%. As can be seen in Fig.~\ref{fig:noise}, this cut removes the majority of noise events, while leaving any events from the iron line untouched.

\begin{figure}[t]
\includegraphics[width=.48\textwidth]{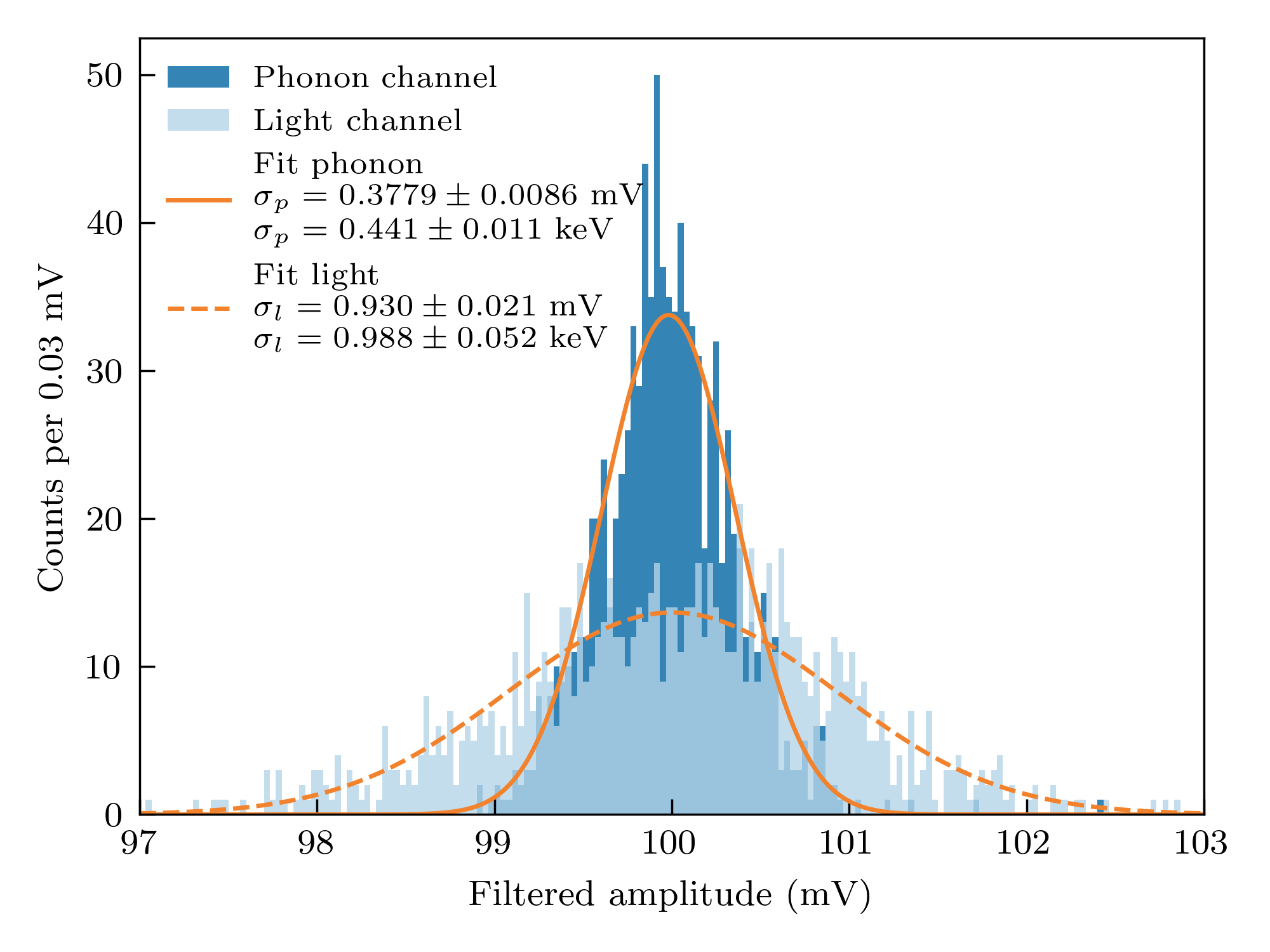}
\caption{\label{fig:resolution} Histogram of the filtered amplitudes of simulated pulses in both phonon (blue) and light channel (light blue), used to determine the detector baseline resolutions. The solid/dashed line is a fit of a Gaussian to the phonon/light channel data, the width of which gives a measure of the baseline resolution.}
\end{figure}

Along with the methodology in the main text, we also provide here Fig. \ref{fig:resolution} to illustrate how the baseline resolution was determined from fitting a Gaussian distribution to the filtered set of simulated events. 

\begin{figure}[t]
\includegraphics[width=.48\textwidth]{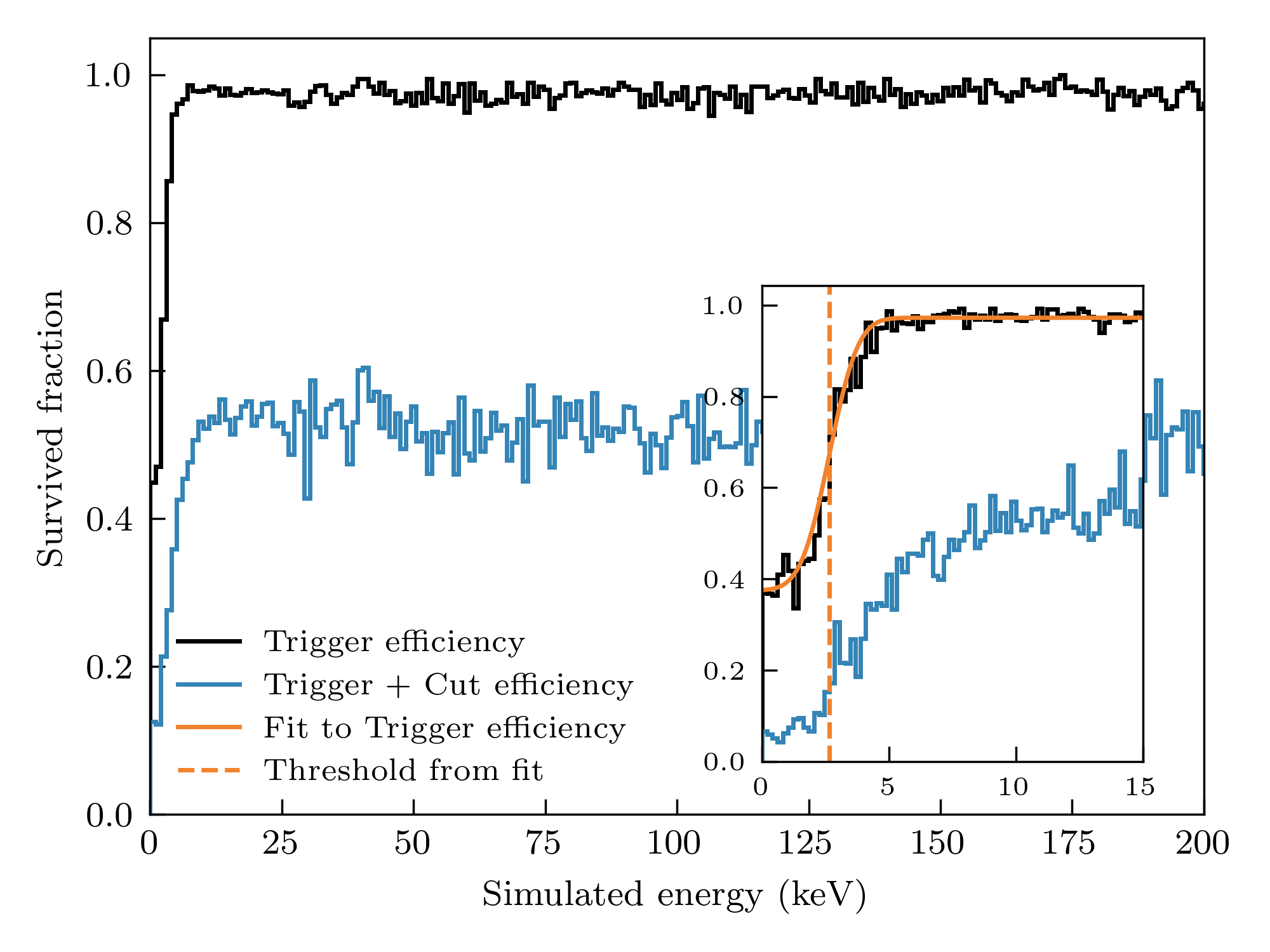}
\caption{\label{fig:efficiency}The black line shows the trigger efficiency, the blue line the trigger and cut efficiency as determined from simulated pulses. In the inset we show a zoom-in to lower energies, showing also an error function fit to the trigger efficiency (solid orange line), which can be used to estimate the detector threshold to $2.656 \pm 0.041\,$keV (dashed orange line).}
\end{figure}

In Fig.~\ref{fig:efficiency} we illustrate the results from determining the trigger and cut efficiencies. We show the the binned fraction of triggered artificial events in the phonon channel, after removing artificial pile-ups, in black, and the fraction of events surviving also the subsequent analysis chain in blue. An extended error function is fitted to the trigger efficiency (orange line)
\begin{align}
    \label{eq:efficiency}
    \text{eff}_\text{trigger} (E) =c \left((1-\epsilon)\times0.5
    \erf{\left(\frac{(E-t_p)}{\sqrt{2}\sigma}\right)}+\epsilon\right),
\end{align}

where $c$, $\epsilon$, $\sigma$ and $t_p$ are free parameters. The fit resulted in a detector threshold of $t_p =$ \unit[2.656 ± 0.041]{keV}.
\subsection{High-level analysis and likelihood fit} 
\label{highlevelanalysis_sup}

For an event with light yield $LY = E_l/E_p \neq 1$, the energy in the phonon channel $E_p$ is not a direct measure of the total energy deposited in the target crystal, as a fraction of the energy is dissipated into scintillation light $E_l$. As described in the main text, we calculate the total deposited energy as $E=\eta E_l + (1-\eta) E_p$ \cite{CRESST-II:2014}. The correct value of $\eta$ is determined by gradually increasing $\eta$, applying the shift to the data, and fitting a double peak function to the iron lines in Fig.~\ref{fig:noise}. The correct value for $\eta$ is then the one, for which the fitted resolution of the iron lines is minimal. Minimizing the width of the calibration peaks is equivalent to correcting the tilt in the energy-light yield plane described in \cite{CRESST-II:2014}. With this method, we estimate a scintillation efficiency $\eta$ of 9.1\% and a resolution of the detector at the position of the iron lines of $0.450 \pm 0.007\,$keV which agrees with the baseline resolution.

\begin{figure*}[ht]
\includegraphics[width=\textwidth]{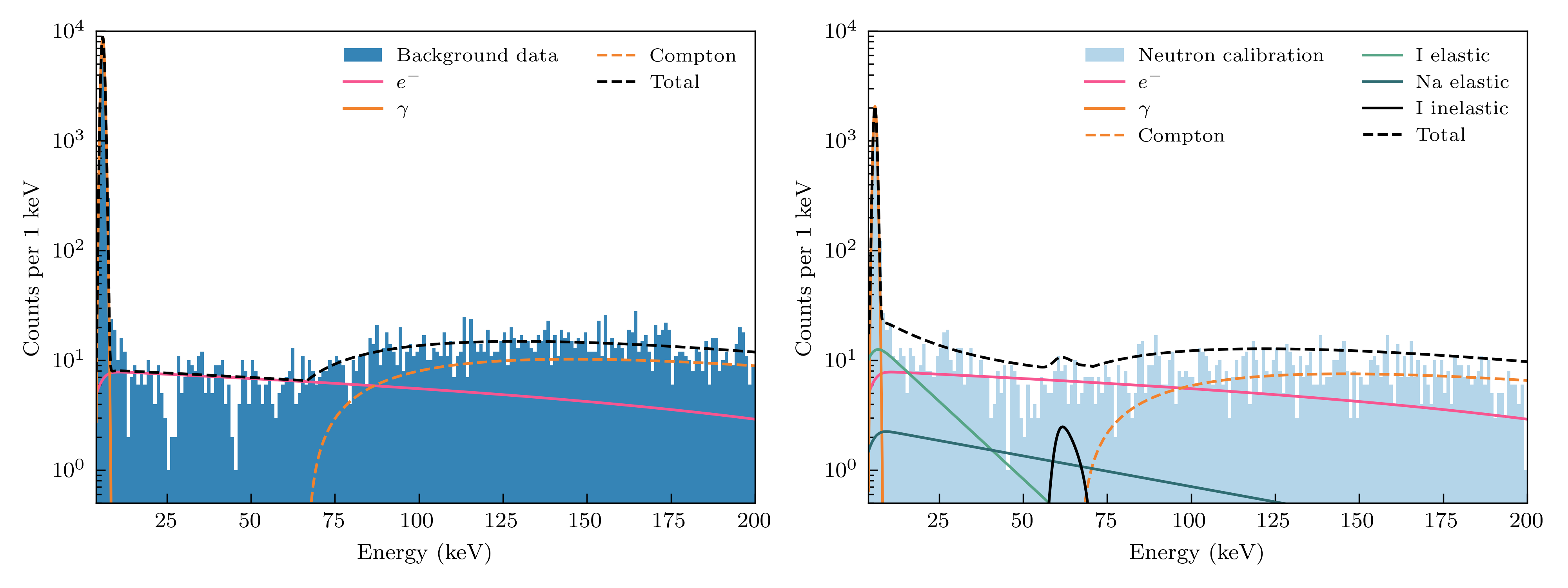}
\caption{\label{fig:spectrum}Energy spectra (total energy, shift corrected) of the ROI for both background (left) and neutron calibration data (right). The plots also show the parametric descriptions of the energy spectra as yielded by the combined likelihood fit.}
\end{figure*}

In this section we also give further details on the parametrization used in the two-dimensional unbinned likelihood-fit to the data and its results. The parametrization of the energy spectra and quenching factors is based on \cite{likelihood}. The minimization of the total likelihood function was performed using \textit{iminuit} \cite{iminuit}, the python implementation of the Minuit framework \cite{minuit}. In addition to the light yield vs.~energy plot in the main text, Fig.~\ref{fig:spectrum} illustrates the performance of the fit in the context of the energy spectra. The selected background model consists of a decreasing flat background, the $^{55}$Fe calibration lines (at \unit[5.89]{keV} and \unit[6.49]{keV}), elastic and inelastic nuclear scattering events, as well as a bump-like description of events caused by $\delta$-electrons created by charged high-energy particles interacting with the surrounding of the detector (labeled "Compton" in Fig.~\ref{fig:spectrum}). The model provides a reasonably good representation of the measured data. We want to note that a full model necessitates further studies and simulations to better understand the possible backgrounds associated with the detector.

One result of the likelihood fit is a description of the energy dependent light quenching of the nuclear band. For the nuclear band $n$ we describe the quenching factor as $QF_n(E)=E_{l,n}(E)/E_{l,e^-}(E)$, where $E_{l,n}(E)$ is the energy dependent mean light of the nuclear recoil band and $E_{l,e^-}(E)$ the mean light for electron recoil. Per definition the mean light yield $E_{l,n}(E)/E_p$ of the electron band is one at the calibration energy, however, we introduce an energy dependent description to account for detector effects  \cite{Strauss_2014}:
\begin{align}
    \label{eq:QF_nuclear}
    E_{l,e^-}(E) &= \left(l_0E+l_1E^2\right)\cdot\left(1-a_{e^-} e^{-E/d_{e^-}}\right), \\
    E_{l,n}(E) &= \left(l_0E+l_1E^2\right)\cdot k_{QF,n}\left(1-a_n e^{-E/d_n}\right)
\end{align}
The variables $l_0,l_1,a_{e^-},d_{e^-},k_{QF,n},a_n,d_n$ are free parameters determined from the fit. Results from the fit, together with the quenching factors at 10 keV, are noted in Tab. \ref{tab:fitresult}. One interesting observation we made in the analysis of this measurement, is that the light yield of the nuclear recoil events increases towards higher deposited energies. This behavior is contrary to observations made in other materials, such as CaWO$_4$ \cite{Strauss_2014}. Using the results from the fit, we then define the acceptance region as the area in the energy - light yield plane between the mean of the sodium and the 99.5\% lower limit of the iodine band. The region is marked in light green in the light yield plot in the main text and contains mostly events leaking from the iron source peak.

We note here, that the exponential description of the energy dependent light yield for the nuclear recoil band is purely phenomenological. To ensure that the parametrization of the nuclear recoil bands has no effect on the DM analysis, we performed the whole high-level analysis chain with various descriptions of the energy dependent light quenching, all leading to comparable limits.

\begin{table}[ht]
\renewcommand{\arraystretch}{1.5}
\begin{tabular}{lrl}
\hline
\textbf{Parameter} &  \multicolumn{2}{c}{\textbf{Fit value}}~~~~ \\\hline 
$l_0$ & $0.8131$&$\pm\,\,0.0026$\\
$l_1$ & $8.98\cdot 10^{-4}$&$ \pm\,\,0.28\cdot 10^{-5}\,\,\text{keV}^{-1}$\\
$a_{e^-}$ & $-0.2279$&$ \pm\,\,0.0043$\\
$d_{e^-}$ & $226$&$ \pm\,\,13	\,\text{keV} $\\
$k_{QF,\text{Na}}$ & $2.864$&$\pm\,\,0.079$\\
$a_\text{Na}$ & $0.9197$&$ \pm\,\,0.0032$\\
$d_\text{Na}$ & $1.91\cdot 10^{3}$&$ \pm\,\,0.12\cdot 10^{3}\,\text{keV} $\\
$k_{QF,\text{I}}$ & $0.1005$&$\pm\,\,0.0041$\\
$a_\text{I}$ & $680$&$ \pm\,\,450$\\
$d_\text{I}$ & $0.071$&$ \pm\,\,0.031\,\text{keV} $\\ \hline
$QF_{\text{Na}}(10\,\text{keV}) $ & $0.2002$&$  \pm\,\,0.0093$\\
$QF_{\text{I}}(10\,\text{keV}) $ & $ 0.0825$&$  \pm\,\,0.0034$\\ \hline
\end{tabular}
\caption{Fit values of the parameters necessary to describe the energy dependent light quenching of the nuclear recoil bands in Eq.~\ref{eq:QF_nuclear} as acquired by the Maximum Likelihood fit. The last two rows state the values of the quenching factors for sodium and iodine at 10 keV total deposited energy.}
\label{tab:fitresult}
\end{table}

\section{Simulation study}\label{simulation}
\begin{figure}[h]
\includegraphics[width=.48\textwidth]{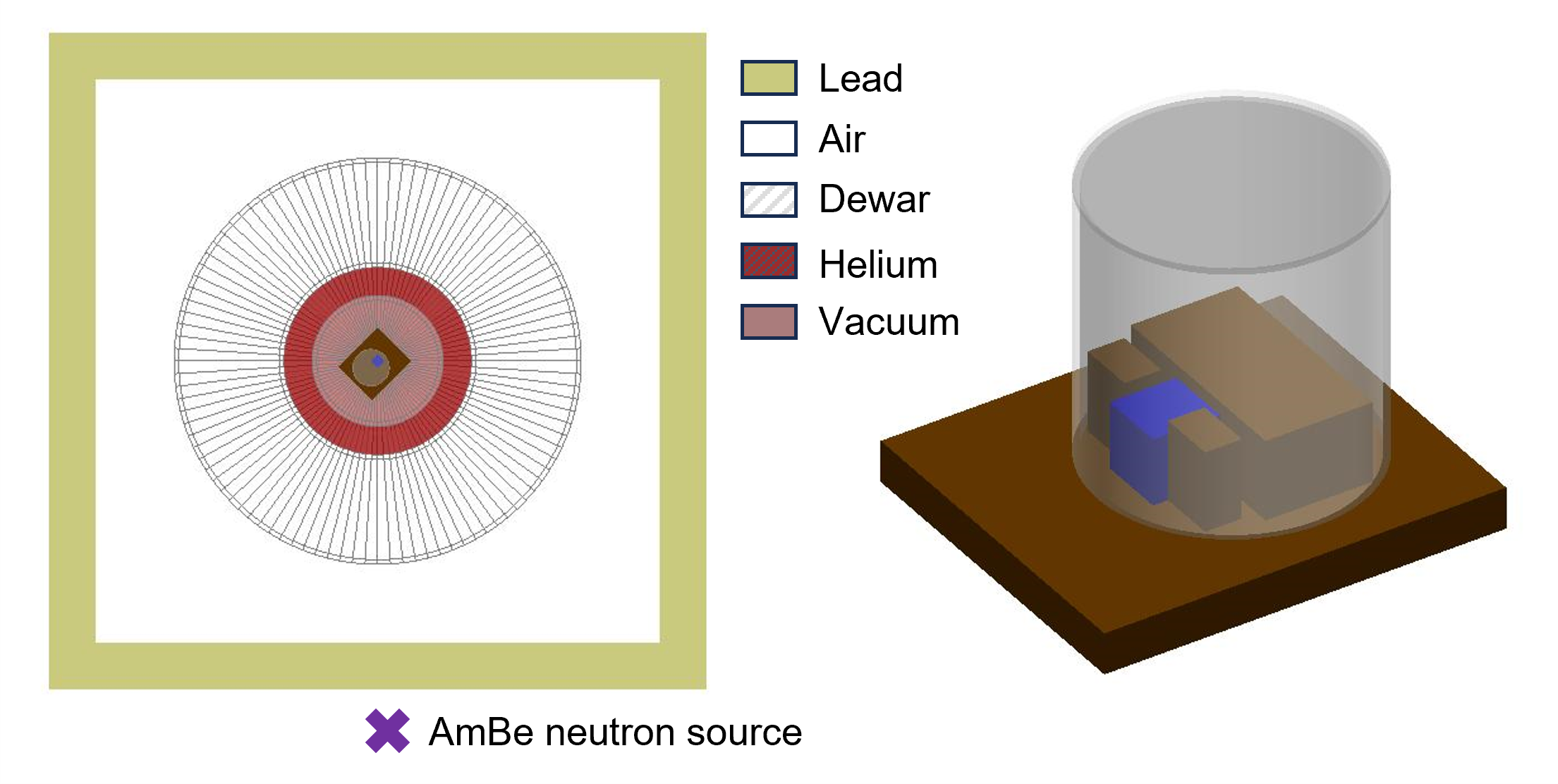}
\caption{\label{fig:geometry}Geometry as simulated in GEANT4. On the left: top view on the refrigerator. On the right: 3D view of the simulated detector module.}
\end{figure}

\begin{figure}[ht]
\includegraphics[width=.48\textwidth]{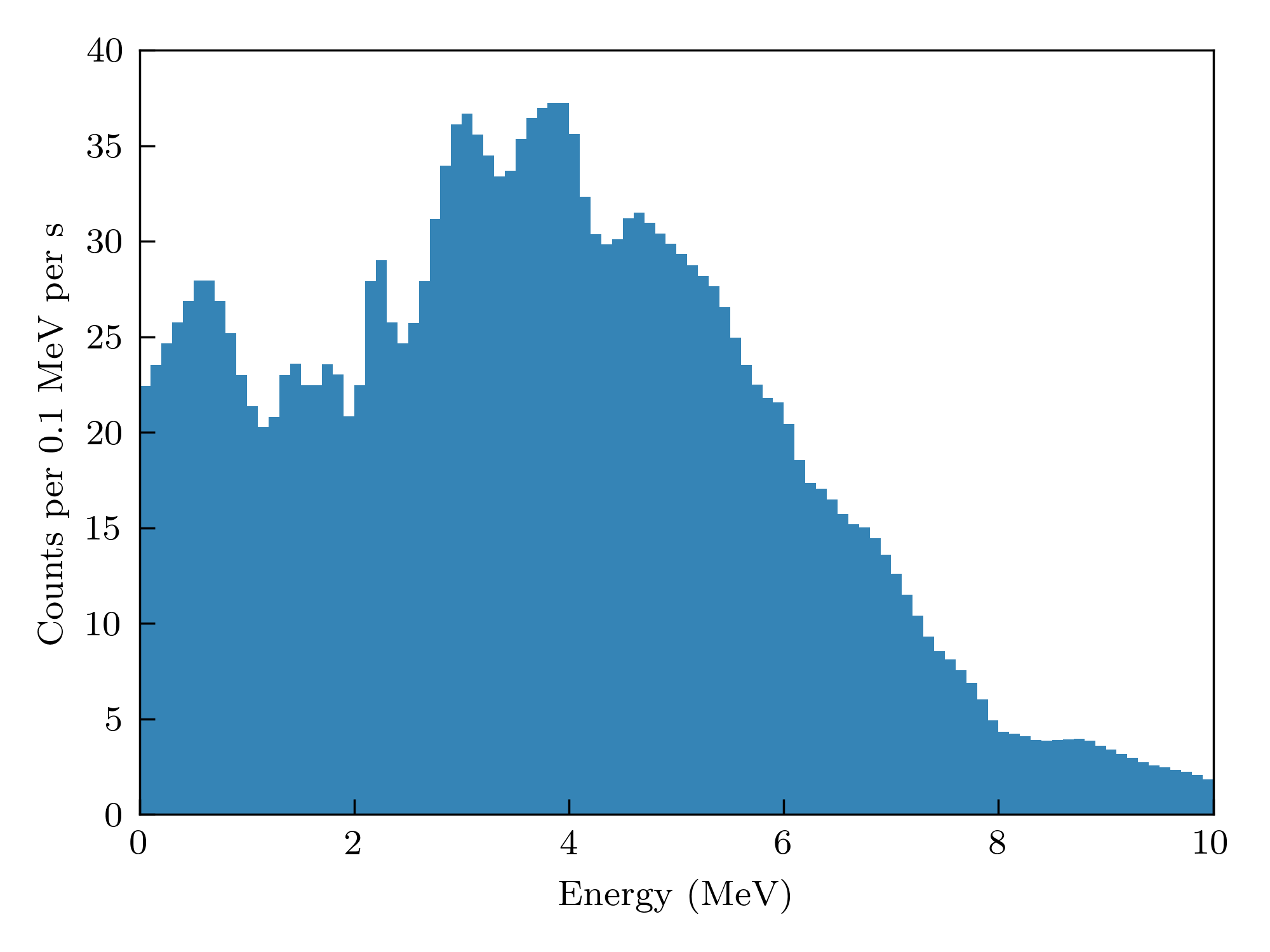}
\caption{\label{fig:sim_spectrum}AmBe spectrum from one million simulated events in an energy region from 0 to \unit[10]{MeV}.}
\end{figure}

To get an estimate of the expected neutron interaction rate in the phonon channel for the measurement with the \textrm{Am}\textrm{Be} source, a GEANT4 (v11.0) based simulation \cite{geant4_1,geant4_2,geant4_3} of the setup was carried out.
The simulated geometry includes the detector module, the vacuum chambers and helium bath of the refrigerator and the external lead shield.
A top view of the setup as implemented in the software is shown in Fig.~\ref{fig:geometry}
An isotropically emitting, point-like neutron source with an activity of \unit[2000]{neutrons/s} was placed next to the external lead shielding, adding  \unit[1]{cm} of polyethylene between.
In Fig.~\ref{fig:sim_spectrum} the spectrum of the simulated neutron source is shown.

\begin{figure}[ht]
\includegraphics[width=.48\textwidth]{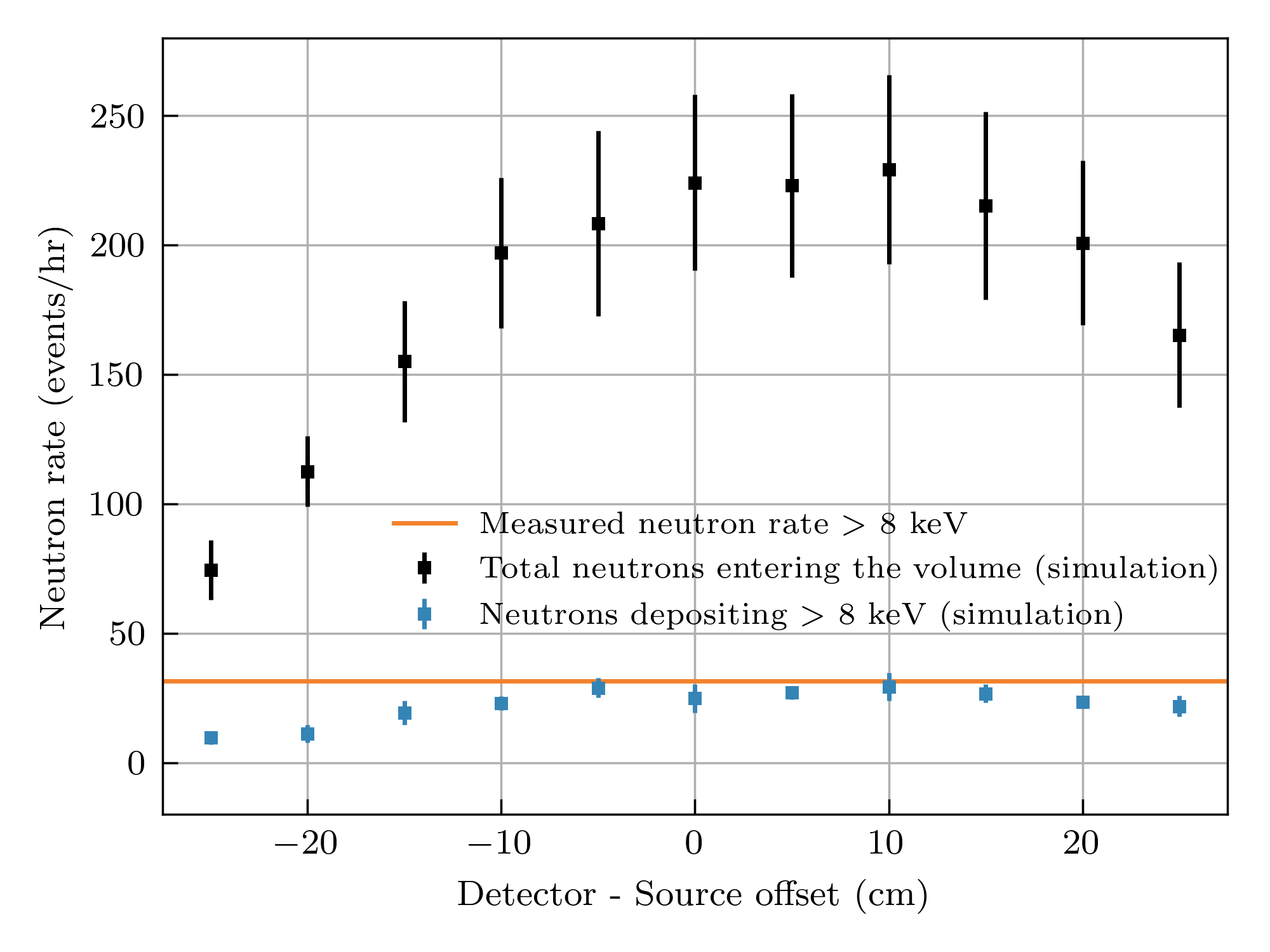}
\caption{\label{fig:sim_rate}Simulated and measured neutron rate in the phonon channel.}
\end{figure}

The height of the source was varied in \unit[5]{cm} increments to the detector plane to estimate the effect of misalignment.
For each position of the neutron source, five simulations with different seeds were computed.
A resulting neutron rate above a given threshold was calculated for each position by computing mean and standard deviation of the simulation results.
To validate the data analysis, the simulated neutron rate can be compared to the measured rate in an energy range which is clearly above threshold (\unit[8]{keV}). For the real data, the neutron rate is calculated by subtracting the rate in the background files, after analysis cuts, from the rate of the neutron calibration data after analysis cuts. The rates from simulation and measurement agree and are displayed in Fig.~\ref{fig:sim_rate}.

\newpage
\bibliography{supplemental_refs.bib}